\documentstyle[prc,aps,preprint]{revtex}

\begin{document}
\draft
\tighten

\title{Targeted free energy perturbation}
       
\author{C. Jarzynski}

\address{Complex Systems, T-13, MS B213 \\
         Los Alamos National Laboratory \\
         Los Alamos, NM 87545 \\
         {\tt chrisj@lanl.gov}\\
         \vskip .2in
         {\rm LAUR-01-2157}\\}

\maketitle

\begin{abstract}
A generalization of the free energy perturbation identity
is derived, and a computational strategy based on this result
is presented.
A simple example illustrates the efficiency
gains that can be achieved with this method.\\
\end{abstract}

The development of efficient methods for the numerical
estimation of free energy differences remains an outstanding
problem in the computational sciences\cite{reviews},
with applications as diverse as rational drug design\cite{drug},
{\it ab initio} prediction of material properties\cite{ackland},
and the study of condensates
in non-perturbative QCD\cite{shafer-shuryak}.
Many schemes for estimating free energy
differences trace their origins to the
{\it perturbation identity}\cite{zwanzig},
\begin{equation}
\label{eq:fep}
\Bigl\langle
e^{-\Delta E/kT}
\Bigr\rangle_A
= e^{-\Delta F/kT}.
\end{equation}
Here, $\Delta F = F_B-F_A$
is the Helmholtz free energy difference
between two equilibrium states of a
system, $A$ and $B$, defined
at a common temperature $T$
but different settings of external
parameters.
The variable ${\bf x}$ (and later ${\bf y}$)
denotes a microstate of the system,
e.g.\ a point in configuration space
or phase space;
$E_A({\bf x})$ and $E_B({\bf x})$
denote the internal energy as a function
of microstate, for the two parameter settings;
and
\begin{equation}
\Delta E({\bf x}) \equiv
E_B({\bf x}) - E_A({\bf x})
\end{equation}
is the energy difference associated
with changing the external parameters from one
setting to the other, while
holding fixed the microstate.
Finally,
$\langle\cdots\rangle_A$ denotes an average
over microstates sampled from the
canonical distribution representing state $A$.

The traditional perturbation approach to estimating
$\Delta F$ is a direct application of Eq.\ref{eq:fep}:
the quantity $\exp(-\Delta E/kT)$ is averaged over
microstates sampled from 
ensemble $A$.\cite{ensembles}
However, this method converges poorly
when there is little overlap in configuration space
between ensembles $A$ and $B$.
Intuitively, this makes sense:
if there is little overlap, then we very slowly
accumulate information about state $B$
by generating microstates typical of state $A$.

The aim of this paper is to present
a generalization of Eq.\ref{eq:fep},
as well as a computational method,
{\it targeted free energy perturbation},
based on this result.
The practitioner of this method must
attempt to construct an invertible transformation ${\cal M}$,
under which ensemble $A$ gets mapped onto an
ensemble $A^\prime$ which overlaps significantly with $B$
(see Eqs.\ref{eq:defaprime}, \ref{eq:approx}).
The more successful this attempt,
the more rapidly the method converges.
Indeed, if $A^\prime$ and $B$ overlap perfectly,
then convergence is immediate!
This strategy thus provides a mechanism
for taking advantage of prior knowledge about states
$A$ and $B$ (used to construct the mapping ${\cal M}$)
in order to speed up the estimation
of $\Delta F$.

While the central result and method derived below
are new, the use of invertible mappings
to enhance the efficiency of free energy
calculations has precedents.
For simple displacements,
${\bf x}\rightarrow{\bf x}+{\bf d}$,
the method proposed herein is closely related to one
developed years ago by Voter\cite{voter},
for energy functions $E_A$ and $E_B$ which
resemble one another apart from a spatial translation.
Bruce {\it et al}\cite{lattice-switch}
have proposed the use of invertible transformations
as collective Monte Carlo moves --
``lattice switches'' -- to enable the sampling of
disparate regions of configuration space.
Finally, our method is similar in
spirit to the metric scaling scheme
developed by Miller and Reinhardt\cite{mr},
whereby one attempts to ``guide'' the system in question
through a continuous sequence of equilibrium
states, by 
dynamically, and linearly, distorting the space in
which the constituent particles evolve.

We now derive our central result,
Eq.\ref{eq:iden} below.

Consider an invertible transformation of
configuration space onto itself:
\begin{equation}
{\cal M}: {\bf x} \rightarrow {\bf y}({\bf x}).
\end{equation}
This might be a displacement
as in Ref.\cite{voter}, or perhaps
a scaling transformation as in Ref.\cite{mr},
or it might be a considerably more complicated,
nonlinear mapping.
Now imagine microstates
${\bf x}_1,{\bf x}_2,\cdots$
sampled from some (so far, arbitrary) 
{\it primary} ensemble,
represented by a distribution $\rho({\bf x})$;
and construct
their images under the transformation:
${\bf y}_1,{\bf y}_2,\cdots$,
where ${\bf y}_n={\cal M}({\bf x}_n)$.
The ${\bf y}$'s are, effectively, sampled from
a {\it secondary} ensemble, which is the
image of the primary ensemble under ${\cal M}$.
This secondary ensemble is represented
by a distribution $\eta(\cdot)$, related 
to the primary distribution $\rho(\cdot)$ by:
\begin{equation}
\label{eq:etarho}
\eta({\bf y}) =
\rho({\bf x})/J({\bf x}),
\end{equation}
where $J({\bf x})=\vert \partial{\bf y}/\partial{\bf x}\vert$
is the Jacobian of the mapping ${\cal M}$.
Here and henceforth, when the variables ${\bf x}$
and ${\bf y}$ appear together, it will
be understood that they are related by
${\bf y}= {\cal M}({\bf x})$.

Next, define a function
\begin{equation}
\label{eq:defPhi}
\Phi({\bf x}) \equiv
E_B({\bf y}) - E_A({\bf x}) - 
kT\,\ln J({\bf x}),
\end{equation}
let the primary ensemble be the canonical
distribution corresponding to state $A$,
\begin{equation}
\label{eq:xA}
\rho({\bf x}) =
{1\over Z_A} e^{-E_A({\bf x})/kT},
\end{equation}
and evaluate the average of
$\exp(-\Phi/kT)$ over points ${\bf x}$
sampled from this ensemble:
\begin{eqnarray}
\Bigl\langle e^{-\Phi/kT} \Bigr\rangle_A
&=& \int d{\bf x}\,\rho({\bf x})\,
e^{-\Phi({\bf x})/kT} \\
&=&
{1\over Z_A} \int d{\bf x}\,
J({\bf x}) e^{-E_B({\bf y})/kT} \\
&=&
{1\over Z_A} \int d{\bf y}\,
e^{-E_B({\bf y})/kT} = {Z_B\over Z_A},
\end{eqnarray}
where $Z_A$ and $Z_B$ are partition functions.
(Note the change in the variable of integration:
$\int J\,d{\bf x}\cdots = \int d{\bf y}\cdots$.)
Invoking the relation
$F = -kT\,\ln Z$, we finally obtain
\begin{equation}
\label{eq:iden}
\fbox{$
\Bigl\langle e^{-\Phi/kT} \Bigr\rangle_A
= e^{-\Delta F/kT}
$}.
\end{equation}
Let us now turn our attention to the application
of this result to the problem of estimating
$\Delta F$.

Eq.\ref{eq:iden} generalizes
of the free energy perturbation identity, reducing to the latter
in the special case ${\cal M}:{\bf x}\rightarrow{\bf x}$.
However, Eq.\ref{eq:iden} is valid for
arbitrary invertible transformations ${\cal M}$.
{\it It is plausible that one can take advantage 
of this generality to enhance the efficiency 
of computing $\Delta F$}.
That is, there may exist
mappings ${\cal M}$ for which the average of 
$\exp(-\Phi/kT)$ converges more rapidly than
the average of $\exp(-\Delta E/kT)$.

To investigate this possibility,
consider $p(\phi\vert {\cal M})$,
the distribution of values of
$\phi=\Phi({\bf x})$,
for ${\bf x}$ sampled from $A$ (Eq.\ref{eq:xA}).
Eq.\ref{eq:iden} asserts that
\begin{equation}
\label{eq:altiden}
\int d\phi\,p(\phi\vert {\cal M})\,e^{-\phi/kT}
= e^{-\Delta F/kT},
\end{equation}
for any choice of ${\cal M}$.
In practice, we
estimate $\Delta F$ by averaging
$\exp(-\Phi/kT)$ over a finite
number of sampled microstates
${\bf x}_1,{\bf x}_2,\cdots,{\bf x}_N$:
\begin{equation}
\label{eq:dFest}
{1\over N}\sum_{n=1}^N
e^{-\phi_n/kT} \approx e^{-\Delta F/kT},
\end{equation}
where $\phi_n\equiv\Phi({\bf x}_n)$.
This approximation becomes an equality as
$N\rightarrow\infty$, but the rate of convergence
depends strongly on the choice of ${\cal M}$;
roughly speaking, the narrower the distribution
$p(\phi\vert {\cal M})$, the faster the convergence.
Therefore we are faced with the practical question,
how do we choose ${\cal M}$ so as to maximize the rate
of convergence of the left side of Eq.\ref{eq:dFest}?

Recall that poor
convergence of the usual perturbation method 
is a symptom of too little overlap between
$A$ and $B$ in configuration space: 
we then learn little about $B$
when sampling from $A$. 
Now note that, when generating the sequence of
$\phi_n$'s, we are effectively harvesting
information from {\it two} ensembles;
the points ${\bf x}$ sample the
primary ensemble $A$, while the points
${\bf y}$ sample
the secondary ensemble, $A^\prime$,
which is the image of $A$ under the transformation ${\cal M}$:
\begin{equation}
\label{eq:defaprime}
A \stackrel{{\cal M}}{\longrightarrow} A^\prime.
\end{equation}
Intuition suggests that, since the primary
ensemble represents state $A$ (by construction),
we ought to attempt to maximize the overlap
between the secondary ensemble and state $B$,
\begin{equation}
\label{eq:approx}
A^\prime \approx B,
\end{equation}
so as to speedily gain information about
both of the equilibrium states ($A$ and $B$)
which interest us.

Pursuing this line of intuition, let us
first consider the extreme case, 
and define a {\it perfect} transformation
${\cal M}^*$ to be one under which
$A$ maps exactly onto $B$:
\begin{equation}
\label{eq:perfect}
\rho({\bf x}) = {1\over Z_A}
e^{-E_A({\bf x})/kT}
\quad
\stackrel{{\cal M}^*}{\longrightarrow}
\quad
\eta({\bf y}) = {1\over Z_B}
e^{-E_B({\bf y})/kT}.
\end{equation}
By Eq.\ref{eq:etarho}, this implies
\begin{equation}
F_B-E_B({\bf y}) =
F_A-E_A({\bf x}) -kT\ln J({\bf x}),
\end{equation}
in other words $\Phi({\bf x}) = \Delta F$
{\it for all} ${\bf x}$.
Hence, we have a maximally narrow distribution of
values of $\phi$:
\begin{equation}
\label{eq:delta}
p(\phi\vert {\cal M}^*) = \delta(\phi - \Delta F).
\end{equation}
Thus,
{\it the convergence of Eq.\ref{eq:dFest}
is immediate if the transformation is perfect}:
$\Phi({\bf x})=\Delta F$ for every sampled ${\bf x}$.

Unfortunately, constructing a perfect transformation
is likely to be much more difficult than the original
problem of computing $\Delta F$.
However, it stands to reason that if
$p(\phi\vert {\cal M})$ is a delta-function when
$A^\prime=B$,
then it will remain narrow when
$A^\prime\approx B$.
Eq.\ref{eq:delta} thus gives credence 
to our earlier intuition (Eq.\ref{eq:approx}):
we ought indeed to look for a
transformation under which $A^\prime$
enjoys good overlap with $B$.
A close resemblance between $A^\prime$ and $B$
implies a narrow distribution of $\phi$'s,
which in turn implies rapid convergence of our
estimate of $\Delta F$.

Let us summarize what has been stated to this point.
Eq.\ref{eq:iden} suggests
a method of estimating $\Delta F$:
microstates ${\bf x}_n$ are sampled from the
canonical ensemble $A$; the value
$\phi_n=\Phi({\bf x}_n)$ is computed for each
sampled microstate;
and the estimator
$X_N\equiv(1/N)\sum_n\exp(-\phi_n/kT)$
converges to $\exp(-\Delta F/kT)$ as $N\rightarrow\infty$.
Two ingredients of this scheme are:
an invertible mapping ${\cal M}$, and the image
$A^\prime$ of the canonical ensemble $A$ under ${\cal M}$.
If $A^\prime$ coincides with $B$,
then the method converges immediately.
Hence, if we choose a transformation
${\cal M}$ which significantly improves the overlap with $B$,
without necessarily being ``perfect'', then $X_N$ ought
to converge more rapidly with $N$ than the traditional
free energy perturbation estimator,
$X_N^{\rm FEP}\equiv(1/N)\sum_n\exp(-\Delta E_n/kT)$.
We will refer to this method as {\it targeted
free energy perturbation}, since its successful
implementation requires finding a transformation
${\cal M}$ for which the secondary ensemble $A^\prime$ comes
reasonably close to ``hitting'' the {\it target ensemble}, $B$.

Several potential extensions of targeted free energy perturbation,
to be developed more fully elsewhere, suggest themselves.
First, for a parameter-dependent energy function
$E({\bf x},\lambda)$, the application of Eq.\ref{eq:iden},
to states $A$ and $B$ defined by infinitesimally different
values of $\lambda$, leads to the identity
\begin{equation}
\label{eq:modTI}
{\partial F\over\partial\lambda} = 
\Biggl\langle
{\partial E\over\partial\lambda} +
{\bf u}\cdot\nabla E
- kT\,\nabla\cdot {\bf u}
\Biggr\rangle_\lambda,
\end{equation}
where ${\bf u}({\bf x})$ is an arbitrary differentiable
vector field of bounded magnitude\cite{derive_modTI}
While this result reduces to the widely used
thermodynamic integration identity \cite{kirkwood}
for ${\bf u}={\bf 0}$,
other choices of ${\bf u}$ might accelerate
convergence of the average.
It is also straightforward to incorporate Eq.\ref{eq:iden} into
the umbrella sampling\cite{tv}
and overlapping distributions\cite{bennett} methods.
Finally, a {\it nonlinear} metric scaling scheme 
-- with particular potential for enhancing the efficiency
of free energy calculations based on
steered molecular dynamics\cite{smd} --
might result from combining the approach of the present
paper with that of Ref.\cite{mr}.

The utility of targeted free energy perturbation depends 
critically on our ability to construct a mapping ${\cal M}$
appropriate to the problem at hand.
While intuition will in some cases reliably suggest
a candidate, in others it may be very difficult 
or computationally expensive to devise a mapping
which improves the overlap with ensemble $B$.
In the case of {\it non-diffusive} systems, however, a promising
and quite general strategy exists.\cite{ls_harmonic}
For a quasi-rigid system such as a large molecule,
the canonical ensemble occupies a strongly localized 
region of configuration space
(assuming that the translational and rotational
degrees of freedom of the entire molecule have been
integrated out, or else pinned down by a
constraining potential).
Given two such molecules, $A$ and $B$ --
alchemically different,
hence represented by different energy functions\cite{drug2} --
we can roughly approximate the associated canonical ensembles
by Gaussian distributions in the many-dimensional
configuration space.\cite{gaussian}
A reasonable candidate for ${\cal M}$ is then
the linear transformation which converts one of these
Gaussians into the other.
Even if the Gaussian approximation is quite crude,
the mapping thus constructed
is likely to result in a significantly improved
overlap between $A^\prime$ and $B$
(relative to that between $A$ and $B$).

We end this paper with numerical results illustrating
the targeted free energy perturbation method.
The setting is the expansion of a cavity in a fluid.
While the aim here is simply a comparison between
methods, 
it bears mention that recent years have seen renewed 
theoretical interest in the problem of cavity
formation in fluids\cite{hydro},
both as a fundamental problem in physical chemistry,
and because of the role played by hydrophobicity in
determining and stabilizing protein structure.

Consider $n_p$ point molecules confined within
a cubic container of volume $L^3$,
but excluded from a spherical cavity of radius
$R$ located at the center (${\bf r}={\bf 0}$) of the container.
Assume periodic boundary conditions and
a pairwise interaction $V_{\rm int}({\bf r}_i,{\bf r}_j)$
between molecules.
We can write the energy function for
such a fluid as:
\begin{equation}
E({\bf x};R) =
\Theta({\bf x};R) +
\sum_{i<j} V_{\rm int}({\bf r}_i,{\bf r}_j).
\end{equation}
Here, ${\bf x}=({\bf r}_1,\cdots,{\bf r}_{n_p})$
specifies the microstate, and
$\Theta({\bf x};R)$ is either 0
(if all $r_i>R$) or $+\infty$ (otherwise),
thus enforcing the exclusion of molecules from
the spherical cavity.
Treating the cavity radius $R$ as an external parameter,
let us choose two values,
$R_A$ and $R_B$, satisfying $0<R_A<R_B<L/2$,
and let $A$ and $B$ denote the corresponding equilibrium 
states (canonical ensembles), at a given temperature $T$.
We want to compute the associated free energy difference,
$\Delta F = F_B-F_A$.
Physically, this is the reversible, isothermal work
required to expand the cavity radius from $R_A$ to $R_B$.

The quantity $\exp(-\Delta F/kT)$ is equal to
the probability $P$ that, given a microstate ${\bf x}$
sampled from ensemble $A$, the region
$R_A<r\le R_B$ will be devoid of molecules.
This can be viewed as a consequence of Eq.\ref{eq:fep},
noting that $\Delta E$ is equal to either
0 or $+\infty$, depending on whether or not this region
is vacant.
The application of the traditional perturbation
method amounts to evaluating this probability
by straight sampling.

Let us now try to construct a transformation
${\cal M}$ which improves the efficiency of estimating 
$P=\exp(-\Delta F/kT)$.
With the traditional method,
poor convergence arises if $P\ll 1$,
i.e.\ if for nearly every microstate ${\bf x}\in A$,
there will be molecules located in the region $R_A<r\le R_B$.
Therefore let us choose ${\cal M}$ so as to vacate this region.
A candidate transformation\cite{geometry},
acting on each particle independently, is:
\begin{equation}
{\bf r}_i \rightarrow g(r_i)\,{\bf r}_i
\qquad,\qquad i=1,\cdots,n_p,
\end{equation}
where
\begin{equation}
g(r) =
\Biggl[
1 + 
{(R_B^3-R_A^3)(L^3-8r^3)\over
(L^3-8R_A^3) r^3}
\Biggr]^{1/3}
\end{equation}
if $R_A<r\le L/2$,
and $g(r)=1$ otherwise.
Under this transformation, the region of
space defined by $R_A<r\le L/2$ gets uniformly
compressed into the region $R_B<r\le L/2$.
For any ${\bf x}$ sampled from $A$,
the quantity $E_B({\bf y})-E_A({\bf x})$ is just the change
in the total interaction energy ($\sum V_{\rm int}$)
resulting from this compression, and
\begin{equation}
J({\bf x}) = [(L^3-8R_B^3)/(L^3-8R_A^3)]^{\nu({\bf x})},
\end{equation}
where $\nu({\bf x})$ is the number of molecules
in the region $R_A<r\le L/2$.

We simulated 125 molecules inside a container
of sides $L=22.28 \,{\rm\AA}$, at $T=300\,{\rm K}$.
A Lennard-Jones interaction between molecules was used,
with parameters corresponding to argon\cite{reid}
($\sigma=3.542 {\rm\AA}$, 
$\epsilon = 0.1854 {\rm kcal/mol}$).
The values of $R_A$ and $R_B$ were taken to be
$9.209\,{\rm\AA}$ and $9.386\,{\rm\AA}$, respectively.

Sampling from ensemble $A$ was achieved
with the Metropolis algorithm.
Three independent runs were carried out,
each consisting of 500 initial relaxation sweeps
followed by $2\times 10^5$ production sweeps.
These runs were used to estimate
$P=\exp(-\Delta F/kT)$, using both
the traditional perturbation approach
(i.e.\ observing the frequency with which the region
$R_A<r\le R_B$ is spontaneously vacant)
and targeted perturbation (Eq.\ref{eq:dFest}).
In Fig.\ref{fig:results}, 
each red curve shows the traditional perturbation
estimate of $P$ for a single run, accumulating
as a function of number of production sweeps, $N$
(plotted in increments of $\Delta N=1000$).
The blue curves show the targeted
perturbation estimates for the same runs.
It is evident that the latter converge much faster
than the former.
Combining the data from all three runs, the two
methods yield the estimates
$P_{\rm trad}^{\rm est} = (4.83 \pm 0.49)\times 10^{-4}$
and
$P_{\rm targ}^{\rm est} = (5.81 \pm 0.05)\times 10^{-4}$.
The error bars are $1\sigma$, and their
ratio suggests that efficiency in this case is improved
by about two orders of magnitude by using targeted
(rather than traditional) free energy perturbation.

It is a pleasure to acknowledge stimulating discussions
and correspondence with Graeme Ackland, Alastair Bruce,
Angel Garcia, Gabriel Istrate, Lawrence Pratt,
and Nigel Wilding.
This research is supported by the Department of Energy,
under contract W-7405-ENG-36.

\begin{figure}
\caption{
Traditional and targeted free energy perturbation
estimates of $P=\exp(-\Delta F/kT)$, as a function
of number of MC sweeps.}
\label{fig:results}
\end{figure}


\begin{references}
\bibitem{drug2}
The problem of estimating $\Delta F$
between alchemically different molecules or
molecular complexes is a central problem of
rational drug design;
see Ref.\cite{drug}.
\bibitem{geometry}
For more complicated geometries, analogous
mappings could be constructed numerically
at relatively low cost.
\end{references}
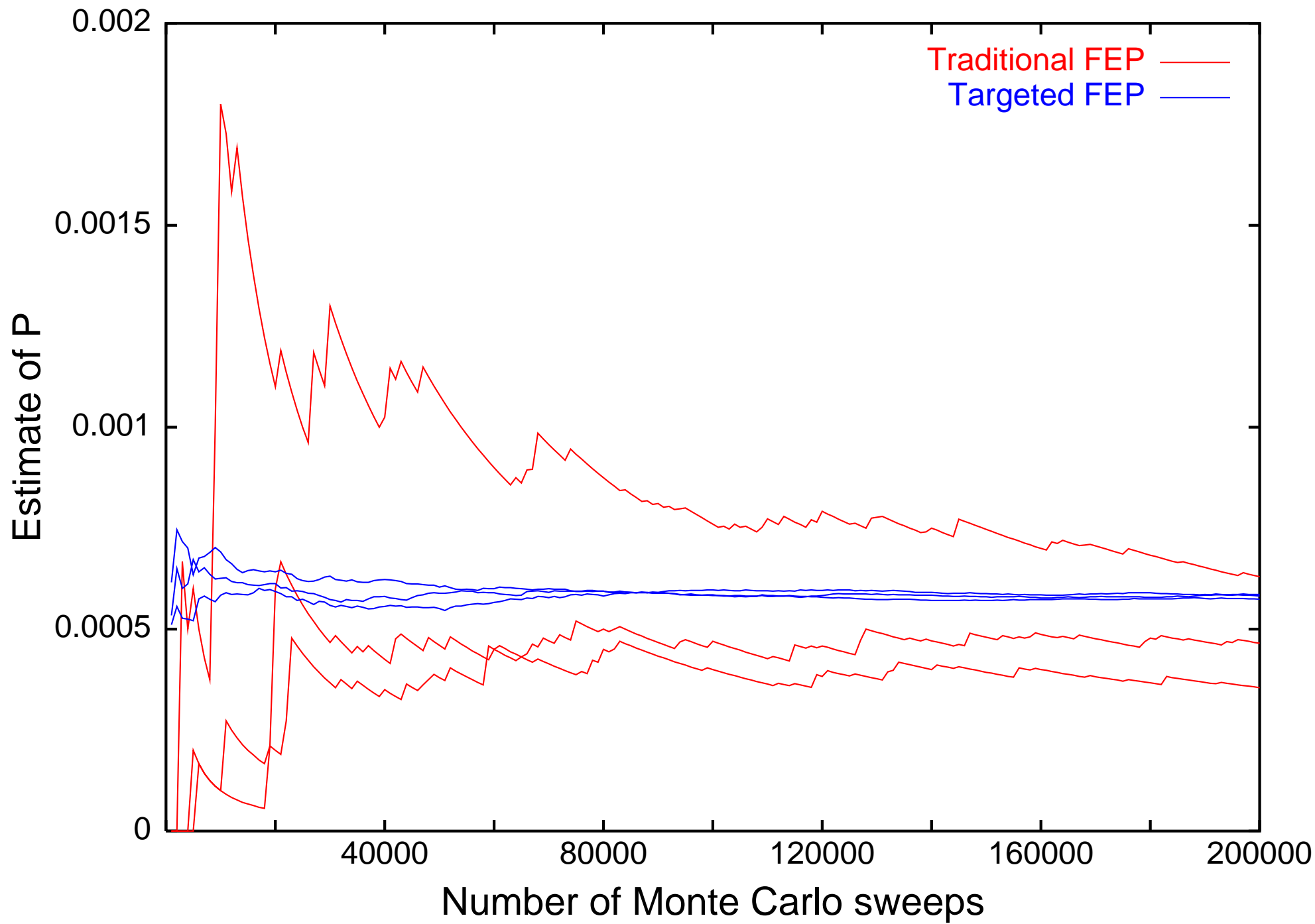
\end{document}